\documentclass[aps,pra,10pt,onecolumn,amsmath,amssymb]{revtex4-2}

\usepackage{mathtools}
\usepackage{graphicx}
\usepackage{xcolor}

\usepackage{placeins}
\usepackage{float}
\usepackage{flafter}

\allowdisplaybreaks

\begin{document}

\title{Structural Perturbations and Full Field Evolution in Tapered Spun Ring-Core Fiber Amplifiers}

\author{Hassan Asgharzadeh B.}
\email{hassan.asgharzadehbastehdimi@tuni.fi}
\affiliation{Faculty of Engineering and Natural Sciences, Tampere University, FI-33014 Tampere, Finland}

\author{Regina Gumenyuk}
\affiliation{Faculty of Engineering and Natural Sciences, Tampere University, FI-33014 Tampere, Finland}

\author{Marco Ornigotti}
\affiliation{Faculty of Engineering and Natural Sciences, Tampere University, FI-33014 Tampere, Finland}

\date{\today}

\begin{abstract}
\noindent We theoretically investigate the complex full-vector modal field evolution in tapered spun ring-core fiber amplifiers subject to structural perturbations, utilizing a combined coupled-mode theory and rate-equation model. To provide a comprehensive characterization, we introduce a normalized complex modal-state fidelity factor that simultaneously incorporates intensity, phase and polarization into a single quantitative measure. We demonstrate that the twist pitch of the structure can be purposely tuned to either suppress or leverage the effect of specific geometrical perturbations. These findings provide insight into the physical conditions governing modal stability, establishing a robust mechanism for highly efficient in-fiber OAM mode conversion. 
\end{abstract}

\maketitle

%%%%%%%%%%%%%%%%%%%%%%%%%%  body  %%%%%%%%%%%%%%%%%%%%%%%%%%
\section{Introduction}
The exponential growth of global data traffic has driven optical communications to its capacity limits, motivating the exploration of spatial-division multiplexing to use new spatial degrees of freedom of light \cite{gong2019optical}. In spatial domain, orbital angular momentum ($\text{OAM}$) modes have established as an exceptional framework for terabit-scale data transmission\cite{bozinovic2013terabit, zhao2025recent}. While transmitting many spatial modes over long distances in multimode fibers offers massive capacity scaling, there is profound challenges arises from modal crosstalk and polarization mixing \cite{bozinovic2013terabit,lee2019analysis}. To overcome these issues, ring-core fibers ($\text{RCF}$) have been introduced to facilitate stable propagation of $\text{OAM}$ modes \cite{ma2021generation, bozinovic2013terabit, wang2020low} by enhancing mode isolation\cite{bozinovic2013terabit,asgharzadeh2026stability}. However, unavoidable fabrication defects, such as fiber ellipticity and intrinsic birefringence, break the cylindrical symmetry of the waveguide triggering significant modal crosstalk\cite{yue2012mode}. Consequently, these perturbations degrade the modal purity in standard RCFs. Applying a continuous helical twist to the fiber structure provides a practical mechanism to control these effects \cite{przhiyalkovsky2017polarization, barlow1981birefringence, ibarra2021numerical}.
Scaling these passive propagation concepts into active, high-power regimes such as fiber amplifiers introduces additional complexities regarding the gain dynamics and transverse modal instability \cite{zalesskaia2026high}. In this context, tapered spun fiber architectures offer a promising platform for high-power amplification, combining enhanced pump absorption with reduced thermal and nonlinear effects\cite{fedotov2021large}.
%To suppress thermal and nonlinear issues, and optimize pump absorption tapered spun fibers are widely used.

Realizing the potential of these architectures for OAM amplification requires a deeper understanding of how structural twisting and symmetry-breaking perturbations jointly influence modal evolution in the presence of optical gain. In this work, we investigate these effects using a theoretical framework combining coupled-mode theory (CMT) with steady-state rate equations (REs) to evaluate modal stability in tapered spun ring-core fiber amplifiers (TSRCFAs).
%Despite theses advancements, to the best of our knowledge, a detailed theoretical understanding of how specific symmetry-breaking perturbations interact with structural twisting in active tapered multimode OAM amplifiers is currently lacking. In this paper, we investigate this critical gap by developing a detailed theoretical framework following a combined coupled-mode theory (CMT) and steady state rate-equation (RE) model to evaluate modal instability against geometrical perturbations in tapered spun ring-core fiber amplifiers (TSRCFAs). 
To quantitatively assess variations across the complete modal field, we introduce a normalized complex modal-state fidelity factor that moves beyond simple polarization measures to simultaneously capture changes in phase, intensity and polarization. By systematically investigating the interplay between structural perturbations and fiber chirality, we demonstrate that increasing the spinning rate of the twisted structure stabilizes the guided modes against elliptical perturbations, providing highly stable propagation. We further demonstrate that  precise adjustment of the twist pitch can be used to achieve highly efficient in-active-fiber OAM mode converters. Additionally, we identify a cascaded conversion mechanism that extends this capability to higher-order OAM modes. This study provides a solid physical framework for analyzing structural perturbations and designing robust OAM fiber amplifiers, efficient in-fiber OAM modes converters and long-distance OAM transmission systems. 

The paper is structured as follows: Section~2 introduces the theoretical framework for evaluating modal field evolution under geometrical perturbation in TSRCFAs. Section 3 details the specific fiber amplifier specifications under study and discusses the simulation data. Finally, sect.~4 provides a summary of our main findings.

\section{Theory}
\subsection{Modal Dynamics}

To study the evolution of the intensity, polarization map and phase distributions of guided modes propagating through geometrically perturbed TSRCFAs, we employ the rigorous set of CMEs for spun fibers, as detailed in \cite{asgharzadeh2026optical,snyderLove}. The study is fully conducted in the laboratory frame, where the spun fiber is modeled as a $z$-dependent periodic helical structure with $\Lambda$ and $\Upsilon=2 \pi/\Lambda$ as the twist pitch and twist rate, respectively. The general set of CMEs can be expressed as
% \begin{equation}
    % \frac{d{\widetilde{a}}_j(z)}{dz}=\sum_{q}{{\widetilde{a}}_q\left(z\right)}\left\{i\,\mathbb{K}_{jq}%(z)\,e^{i\left(\beta_q-\beta_j\pm\left\{J_q-J_j\right\}\Upsilon\right)z}\right\}.
   %  \label{eq:cmes}
 %\end{equation}
 
  \begin{equation}
     \frac{d{\widetilde{a}}_j(z)}{dz}=\sum_{q}{{\widetilde{a}}_q\left(z\right)}\left\{i\mathbb{K}_{jq}e^{i\left(\beta_q-\beta_j\pm\left\{J_q-J_j\right\}\Upsilon\right)z}\right\}+\frac{1}{2}g_{j}(z)\widetilde{a}_j(z).
     \label{eq:cmesgain}
 \end{equation}
where the last term on the right-hand-side of Eq.~(\ref{eq:cmesgain}) including modal gain factors, $g_{j}(z)$, extends the set of CMEs derived for passive fibers \cite{asgharzadeh2026optical} to active ones \cite{zalesskaia2026high}, following a combined RE and CMT analysis \cite{ono2023static,lu2012general}. The local population inversion $N_{2}(r,\phi)$ is calculated based on the spatially varying intensities of the pump and signal modes. Because the maximum intensity of $\text{OAM}$ modes shifts radially outward with increasing topological charge \cite{asgharzadeh2026stability}, these modes experience differential radial gain. Therefore, accounting for radial spatial hole burning (RSHB) is essential to accurately evaluate modal dynamics. Parameter $\mathbb{K}_{jq}(z)$ represents the total perturbation-induced coupling coefficient between interacting modes $j$ and $q$, and $z$ expresses the propagation direction. Propagation constants and the complex amplitude of the guided OAM modes are shown by $\beta$ and $\widetilde{a}(z)$, respectively. The self-coupling term $\mathbb{K}_{jj}$ which is a modification to the propagation constants is absorbed in the definition of $\beta$ \cite{snyder1978modes,rao2024chirally}. The total angular momentum (TAM) of circularly polarized (CP) OAM modes is denoted by $J$. As detailed in \cite{asgharzadeh2026optical}, the taper-induced interactions are negligible (in particular, compared to those induced by core offset and ellipticity \cite{asgharzadeh2026stability}) when tapering is adiabatic. Accordingly, these contributions are neglected in the present model. However, because fiber's radii are longitudinally varying, modal fields and propagation constants become $z$-dependent \cite{snyderLove}.

All spatial, polarization, and phase information of different guided modes can be rigorously extracted from the complex modal amplitudes $\widetilde{a}_{j}(z)$  multiplied by the 2D spatial field profiles of their respective eigenmodes. 
\subsection{Intensity, Polarization and Phase}
The total electric field at any point in perturbed fibers can be expressed as the coherent superposition of all guided reference $\text{OAM}$ modes 
\begin{equation}
    \widetilde{\boldsymbol{E}}_{\text{total}}(r, \phi, z)=\sum_{j}\widetilde{a}_{j}(z)\,\boldsymbol{E}_{j}^{t}(r,\phi),
\label{totfield}
\end{equation}
where $\boldsymbol{E}_{j}^{t}(r,\phi)$ represents the transverse modal field of the $j$-th guided CP-OAM mode \cite{asgharzadeh2026optical}. Each guided mode is normalized to carry one Watt of optical power, $\lvert\widetilde{a}_{j}\rvert^{2}=1\,\text{W}$ \cite{snyderLove}. 
%so that the total signal power in the perturbed fiber can be written as $\widetilde{P}_{total}=\sum_{j}\lvert\widetilde{a}_{j}\rvert^{2}$ \cite{snyderLove}.

The full set of Stokes parameters $(S_{0},S_{1},S_{2},S_{3})$ can be then directly calculated from the local total electric field in Eq. \eqref{totfield}, in the usual manner \cite{bornWolf}, and the orientation and ellipticity angles can be then defined as \cite{jackson1962classical}
\begin{align}
\psi(r, \phi) &= \frac{1}{2} \arctan\left(\frac{S_2}{S_1}\right), \\[1ex]
\chi(r, \phi) &= \frac{1}{2} \arcsin\left(\frac{S_3}{S_0}\right),
\end{align}
where $\psi(r, \phi)$ represents the orientation angle of the local polarization ellipses relative to the $x$-axis and $\chi(r, \phi)$ is the ellipticity angle, with $\chi=0$ corresponding to purely linear polarization, and $\chi = \pm 45^\circ$ to purely circular polarization. The first Stokes parameter, $S_{0}$, represents the total intensity distribution.

To characterize and visualize the topological phase information, the spatially resolved phase pattern can be extracted from either the $x$ or $y$ components of the total electric field. Here, without loss of generality, we use the $x$-component, but note that choosing the $y$-component instead will give a similar result, as both field components contain the same amount of information. To do so, we then extract the local phase of the $x$-component of the total field as

%the $x$-component of the total electric field (or the $y$-component that contains the same physical information) as
 \begin{equation}
\begin{aligned}
    \Phi_{\text{x}}(r, \phi, z) = \arg(E_{\text{x}}^{\text{tot}}(r, \phi, z)).
\end{aligned}
\label{CPtotalphase}
\end{equation}
where $E_{\text{x}}^{\text{tot}}(r, \phi, z)$ represents the coherent sum of the $x$-component of the total electric field at a given propagation distance $z$.
%project the total electric field, Eq.~(\ref{totfield}), onto the orthogonal basis of LCP and RCP as

% \begin{equation}
%\left\{
%\begin{aligned}
%    E_{LCP}^{total}(r, \phi, z) &=\sum_{j}\frac{1}{\sqrt{2}}\,(E_{x}^{total}-%iE_{y}^{total}), \\
%     E_{RCP}^{total}(r, \phi, z) &=\sum_{j}\frac{1}{\sqrt{2}}\,(E_{x}^{total}+iE_{y}^{total}).
%\end{aligned}
%\right.
%\label{CPtotalfield}
%\end{equation}

 %\begin{equation}
%\left\{
%\begin{aligned}
   % \Phi_{\text{LCP}}(r, \phi, z) = \arg(E_{\text{LCP}}^{\text{tot}}(r, \phi, z)), \\
   %  \Phi_{\text{RCP}}(r, \phi, z) = \arg(E_{\text{RCP}}^{\text{tot}}(r, \phi, z)).
%\end{aligned}
%\right.
%\label{CPtotalphase}
%\end{equation}
%
\subsection{Perturbation-Induced Interactions}
In the laboratory frame, any intrinsic perturbation within the spun fiber structure, such as core offset and ellipticity, rotates continuously along the propagation axis $z$ at the twist rate $\Upsilon$. 
%However, macro-bending induced by coiling the fiber after drawing is fixed in the lab frame and does not spin.
Therefore, the total coupling matrix $\mathbb{K}_{\text{total}}$ can be expressed as 
%\begin{equation} 
%\mathbb{K}_{\text{total}}(z) = \mathbb{K}_{\text{static}} + \mathbb{K}_{\text{rotating}}\,e^{-i (J_{j} - J_{q}) %\Upsilon z}
%\label{totalcoup}
%\end{equation}

\begin{equation} 
\mathbb{K}_{\text{total}}(z) = \mathbb{K}_{jq}\,e^{-i (J_{j} - J_{q}) \Upsilon z},
\label{totalcoup}
\end{equation}
where the term $e^{-i (J_{j} - J_{q}) \Upsilon z}$ evaluates the relative phase accumulation between any two modes based on their TAM.

In general, the modal coupling coefficient $\mathbb{K}_{jq}$ can be written as \cite{snyderLove,asgharzadeh2026optical}
\begin{equation}
\mathbb{K}_{jq}=\frac{k_0}{4} \left(\frac{\varepsilon_0}{\mu_0}\right)^{1/2}\iint \boldsymbol{E}_j\cdot\left[\boldsymbol{\delta\varepsilon}\right]\cdot\boldsymbol{E}_q^{*}\,r dr d\varphi, 
\label{eq:ccoetw}
\end{equation}
where $\boldsymbol{E}_j$ and $\boldsymbol{E}_q$ represent the full-vector electric field of the non-perturbed modes $j$ and $q$, respectively, $k_0$ is the vacuum wave number,  and $\varepsilon_{0}$ and $\mu_{0}$ are the vacuum permittivity and permeability, respectively. Depending on the physical origin of the perturbation, the permittivity correction relative to the reference fiber, $\left[\boldsymbol{\delta\varepsilon}\right]$, takes different analytical forms, some of which are detailed in Refs.\cite{asgharzadeh2026optical,snyderLove,Shvets_2009,han2017bend,xu2013higher}. The azimuthal integration of the perturbation terms with the modal angular profiles dictates that the coupling coefficients are strictly zero unless TAM is conserved \cite{asgharzadeh2026optical,Shvets_2009,wang2020demonstration,napiorkowski2018role}, i.e.,
\begin{equation}
    J_j - J_q = \pm N,
    \label{selectionrule}
\end{equation}
where $N$ is the order of the perturbation. For example, core offset is a $N=1$ perturbation, while elliptical deformation of the cross section of the fiber creates a $N=2$ perturbation.
%Intrinsic linear birefringence flips the circular polarization state, $\Delta\sigma=\pm2$, resulting in a TAM change of of $\Delta J=\pm 2$. 
Following the $\text{TAM}$ selection rule defined in Eq.~(\ref{selectionrule}), modal power exchange occurs efficiently if the phase matching condition is satisfied. 
%For static perturbations, such as macro-bending, the phase matching condition for two interacting guided modes, $j$ and $q$, can be written as
%\begin{equation}
   % \Delta\beta = \beta_j - \beta_q = 0.
   % \label{simplephasematching}
%\end{equation}
For rotating perturbations, the phase matching condition can be stated as \cite{asgharzadeh2026optical}
\begin{equation}
    \Delta\beta = \beta_j - \beta_q \pm N \Upsilon = 0,
    \label{twistedphasematching}
\end{equation}
where the term $\pm N \Upsilon$ accounts for the twist-induced phase mismatch. A close inspection of Eq.~(\ref{twistedphasematching}) reveals that if the twist rate $\Upsilon$ is sufficiently large, interacting modes become completely out of phase so that the coupling is totally suppressed. This means that the spinning of the fiber can provide a protection against the rotating perturbations.
\begin{table}[htbp]
\centering
\caption{Simulation parameters for the modeled TSRCFA}
\resizebox{0.9\linewidth}{!}{%
\begin{tabular}{lcc | lcc}
\hline
\textbf{Parameter} & \textbf{Symbol} & \textbf{Value} & \textbf{Parameter} & \textbf{Symbol} & \textbf{Value} \\
\hline
Inner ring radius (narrower side) & $R_{1}$ & $2.4\,\mu\text{m}$ & Fiber length & $L$ & $3\,\text{m}$ \\
Outer ring radius (narrower side) & $R_{2}$ & $6.5\,\mu\text{m}$ & Signal wavelength & $\lambda_{s}$ & $1.040\,\mu\text{m}$ \\
Cladding radius (narrower side) & $R_{3}$ & $62.5\,\mu\text{m}$ & Pump wavelength & $\lambda_{p}$ & $0.976\,\mu\text{m}$ \\
Taper wider-to-narrower ratio & -- & $2$ \cite{asgharzadeh2026optical} & Pump power & $P_{p}$ & $10\,\text{W}$ \\
Ring guide numerical aperture & $\text{NA}$ & $0.11$ & Signal power & $P_{s}$ & $0.1\,\text{W}$ \\
Taper shape factor & $b_{f}$ & $1.25$ \cite{asgharzadeh2026stability} & Active ion concentration & $N_{\text{Yb}}$ & $7.9\times 10^{7}\,\mu\text{m}^{-3}$ \\
\hline    
\end{tabular}%
}
\label{tab:sim_param}
\end{table}
%Because the static perturbations do not physically rotate in the lab frame, there is no twist rate parameter in the phase matching condition, Eq.~(\ref{simplephasematching}). Therefore, the spinning of the fiber provides no protection against static perturbations. 
\subsection{Fidelity analysis}
To quantitatively characterize the stability of the injected mode over propagation, we define the complex modal-state fidelity, $\mathcal{F}(z)$, as
\begin{equation}
     \mathcal{F}(z)=\frac{|\sum_{j}\widetilde{a}_{j}^{*}(0)\,\widetilde{a}_{j}(z)|^{2}}{(\sum_{j}|\widetilde{a}_{j}(0)|^{2})\,(\sum_{j}|\widetilde{a}_{j}(z)|^{2})}.
\label{eq:fidelity}
\end{equation}
Fidelity-based measures have been established to characterize the preservation of modal properties under propagation through complex media \cite{zhou2021high, singh2025enhanced}.
Following this general approach, Eq.~(\ref{eq:fidelity}) quantifies the similarity between the initial and propagated modal fields considered here, and characterizes the preservation of the full field (not solely the polarization or phase distribution). According to this definition, any changes in intensity, phase and/or polarization distributions will affect $\mathcal{F}(z)$. This parameter ranges between 0 and 1 so that a fidelity of unity indicates exact preservation of the normalized complex modal state.
\section{Results and Discussion}

\subsection{Simulation parameters}
\begin{figure}[htbp]
\centering
\includegraphics[width=12cm]{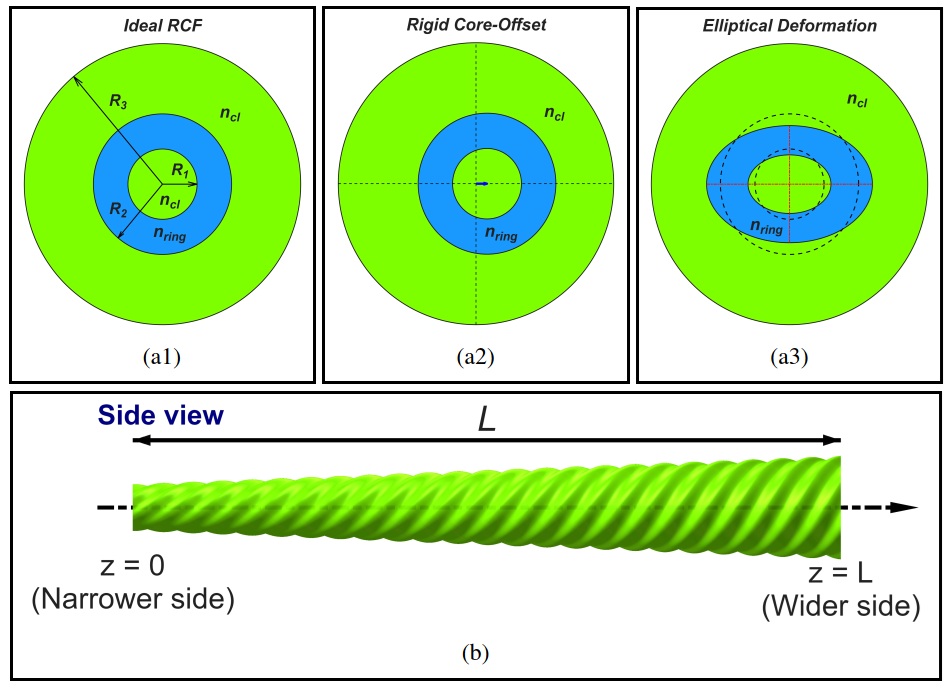}
 \caption{Schematic geometry of a typical TSRCFA. (a1) Ideal RCF cross-section defining the high-index ring-guide ($n_{ring}>n_{cl}$) radii $R_{1,2}$ and cladding radius $R_{3}$. Ring-core cross-sections under (a2) rigid-core-offset and (a3) elliptical perturbations.\\ (b) Longitudinal side view along the propagation axis $z$ illustrating the helical twist and tapered profile over the fiber length $L$.}
    \label{fig:fiber_models}
\end{figure}

To investigate the longitudinal evolution of polarization, intensity, and phase distributions within TSRCFAs, we employ the simulation parameters listed in Table~\ref{tab:sim_param}. Figure \ref{fig:fiber_models}(a1)-(a3) illustrates the cross-sectional geometries of a typical RCF under three distinct structural conditions: and ideal symmetric configuration, a rigid core-offset perturbation, and an elliptical deformation, respectively. The longitudinal side view in Fig.~\ref{fig:fiber_models}(b), schematically shows the physical evolution of the helical twist and the tapering profile along the propagation direction $z$.
The active ring guide is uniformly doped with ytterbium ions ($\text{Yb}^{3+}$). Here, we assume a backward cladding-pumping configuration so that the pump power is injected into the cladding at $z=L$ (the wider side of the taper), while the signal power is launched into the ring guide at $z=0$ (the narrower side of the taper). 
The ring guide supports azimuthal mode order $l=0, 1, \text{and}\, 2$ at the input $z=0$. 
We have not considered higher order modes it this study, as, according to the simulation results, their contribution to modal interactions is negligible under the investigated conditions. However, the study can be readily extended to incorporate them if needed. A similar generalization applies to the pumping configuration. Without loss of generality, we consider a right-handed helical structure for the twisted fibers, corresponding to a positive sign convention for the azimuthal order of the perturbation $N$ \cite{Shvets_2009}.
\subsection{Elliptical deformation}
\begin{figure}[htbp]
\centering
\includegraphics[width=11cm]{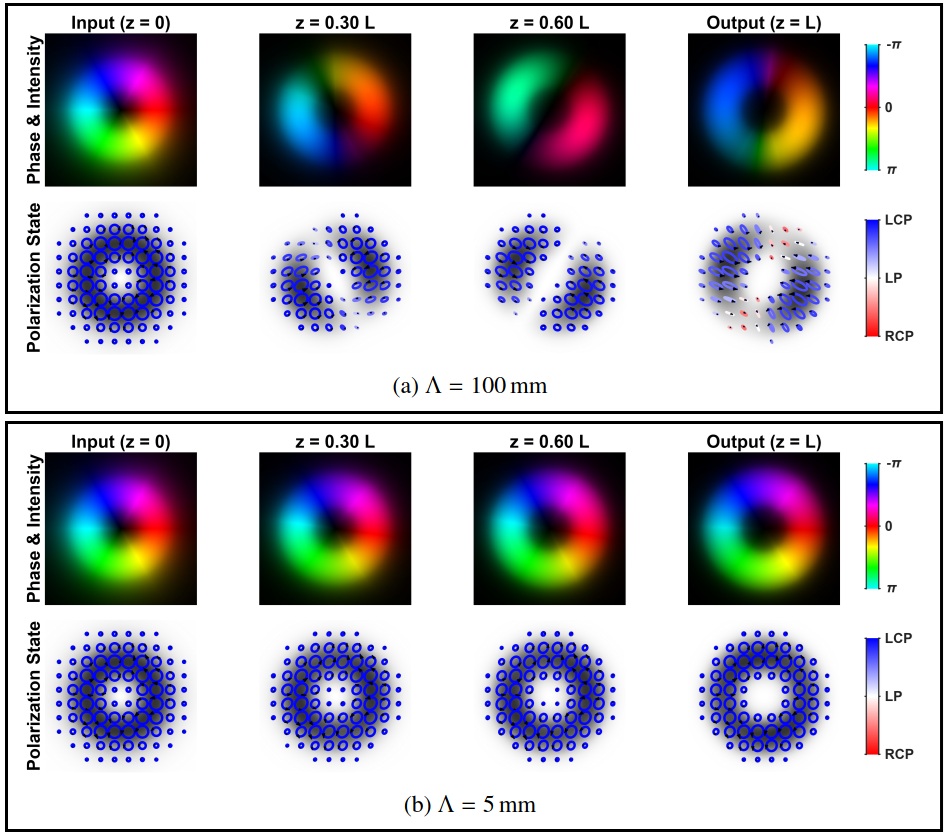}
 \caption{Evolution of the polarization, phase and intensity of an input $\text{LCP OAM}_{11}$ over propagation in the perturbed TSRCFA with twist pitches of (a) $100\,\text{mm}$ and (b) $5\,\text{mm}$, under an elliptical perturbation ($d_{2}=0.3$). Simulation parameters are listed in Table~\ref{tab:sim_param}.}
    \label{fig_ellipticity}
\end{figure}

In this section, we investigate the stability of OAM beams and $\text{TM}$ modes against elliptical deformation ($N=2$ perturbation) of the ring \cite{asgharzadeh2026stability}.
According to Eq.~(\ref{selectionrule}), this perturbation allows coupling between modes differeing by $\Delta J=+2$ in TAM. These modes can belong to modal families with distinct topological charges or belong to the same modal family having orthogonal polarization states. Figure~\ref{fig_ellipticity} shows the evolution of the intensity, phase and polarization distributions of an initially excited $\text{LCP OAM}_{11}$ signal mode over propagation through  TSRCFA introduced in the previous section subjected to an elliptical perturbation amplitude of $d_{2}=0.3$.

The dynamics are compared for twist pitches of $\Lambda=100\,\text{mm}$ (panel (a)) and $\Lambda=5\,\text{mm}$ (panel (b)). A direct comparison reveals that the shorter twist pitch effectively preserves the stability of the injected $\text{LCP OAM}_{11}$ mode. This behavior can be understood by analyzing the phase matching condition, Eq.~(\ref{twistedphasematching}). Because the twist-induced phase mismatch is inversely proportional to the twist pitch, reducing $\Lambda$ to $5\,\text{mm}$ makes the interacting modes completely out of phase resulting in non-efficient modal interactions. This results in robust propagation of $\text{LCP OAM}_{11}$ against the elliptical perturbation. 
While Fig.~\ref{fig_ellipticity} is visually intuitive, the fidelity factor provides a rigorous quantitative measure to analyze the evolution of the complete vector field (polarization, phase and intensity) along the fiber amplifier. For $\Lambda=100\,\text{mm}$ the fidelity factor evaluated at four distinct propagation distances drops significantly as $100\,\%$, $\approx 27\,\%$, $\approx 45\,\%$, and $\approx 52\,\%$, respectively. In contrast, for $\Lambda=5\,\text{mm}$ the fidelity remains approximately untouched as $100\,\%$, $\approx 98\,\%$, $\approx 98\,\%$, and $\approx 99\,\%$, confirming the robust stabilization of the injected mode at the shorter twist pitch. The small fluctuations in the fidelity values originate from the periodic, sinusoidal nature of weak modal interactions.
%\begin{figure}[!t]
%\centering\includegraphics[width=12.5cm]%{Fig1_d2_100mm_twist_OAM11.jpeg}
%\caption{d2-100mm-OAM11}
%\label{d2_100mm_OAM11}
%\end{figure}

\begin{figure}[htbp]
\centering
\includegraphics[width=11cm]{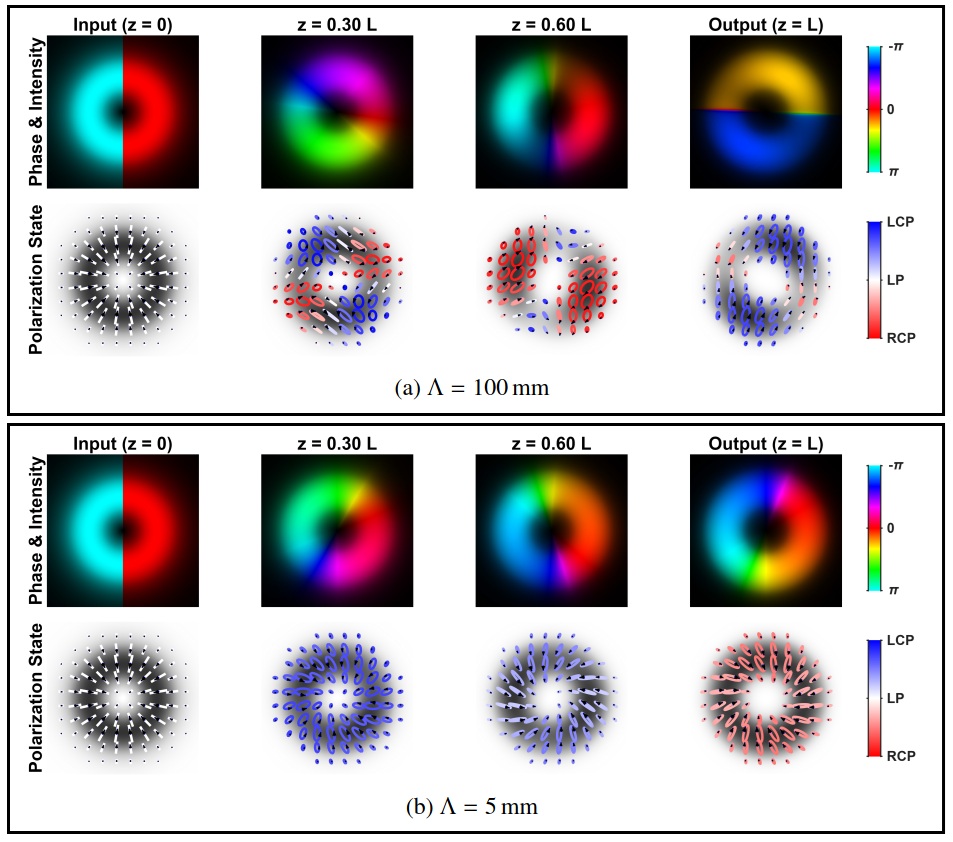}
  \caption{Evolution of the polarization, phase and intensity of an input $\text{TM}_{01}$ over propagation in the perturbed TSRCFA with twist pitches of (a) $100\,\text{mm}$ and (b) $5\,\text{mm}$, under an elliptical perturbation ($d_{2}=0.3$). Simulation parameters are listed in Table~\ref{tab:sim_param}.}
    \label{fig_ellipticity_TM}
\end{figure}

%\begin{figure}[!t]
%\centering\includegraphics[width=12.5cm]%{Fig2_d2_5mm_twist_OAM11.jpeg}
%\caption{d2-100mm-OAM11}
%\label{d2_5mm_OAM11}
%\end{figure}
To extend this analysis to $\text{non-OAM}$ beams, we make a similar analysis for an initially injected $\text{TM}_{01}$ mode (injected at $z=0$) as depicted in Fig.~\ref{fig_ellipticity_TM}.

Analyzing Fig.~\ref{fig_ellipticity_TM} visually, indicates that while reducing the twist pitch to $5\,\text{mm}$ improves the stability of the input $\text{TM}_{01}$, it remains generally more sensitive to the elliptical perturbation compared to $\text{LCP OAM}_{11}$. To assess this observation, we analyze the corresponding fidelity factor. For $\Lambda=100\,\text{mm}$ the fidelity factor drops as $100\,\%$, $\approx 7\,\%$, $\approx 25\,\%$, and $\approx 26\,\%$, while for $\Lambda=5\,\text{mm}$ it varies as $100\,\%$, $\approx 77\,\%$, $\approx 84\,\%$, and $\approx 87\,\%$. A careful comparison of these numbers to those obtained for $\text{LCP OAM}_{11}$ demonstrates the higher sensitivity of $\text{TM}_{01}$ to $N=2$ perturbations.
%\FloatBarrier
\subsection{Core offset}
\begin{figure}[htbp]
\centering
\includegraphics[width=11cm]{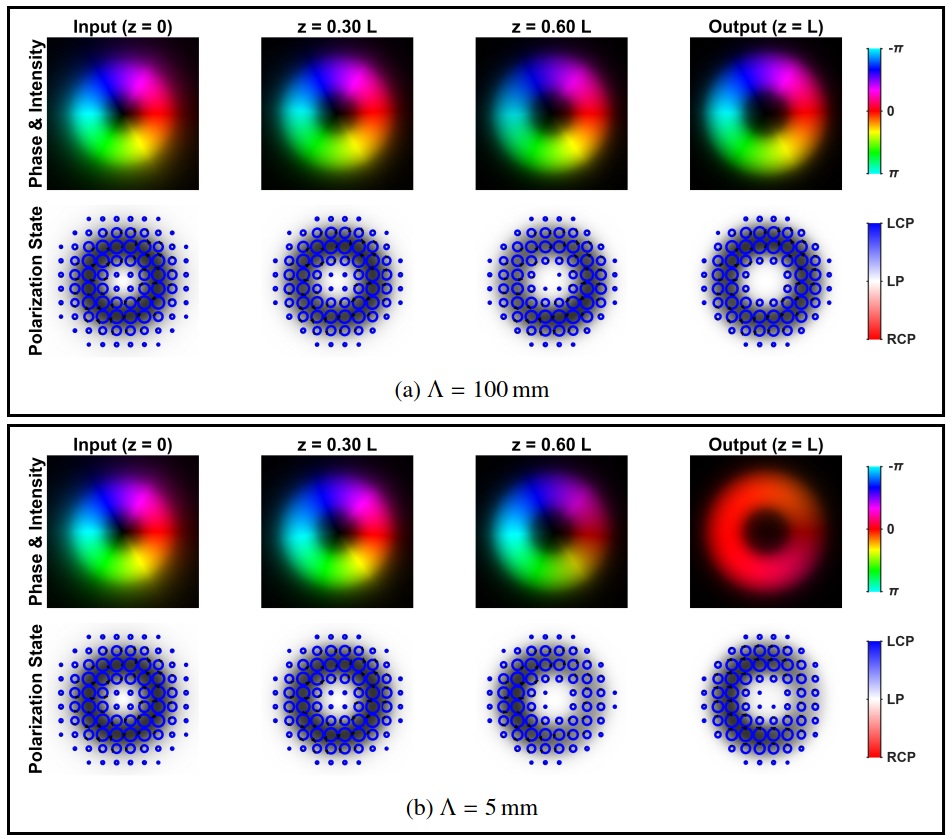}
 \caption{Evolution of the polarization, phase and intensity of an input $\text{LCP OAM}_{11}$ over propagation in the TSRCFA with twist pitches of (a) $100\,\text{mm}$ and (b) $5\,\text{mm}$, under core-offset perturbation ($d_{1}=0.3$). Simulation parameters are listed in Table~\ref{tab:sim_param}.}
    \label{fig_core_offset_OAM11}
\end{figure}

\begin{figure}[htbp]
\centering
\includegraphics[width=12cm]{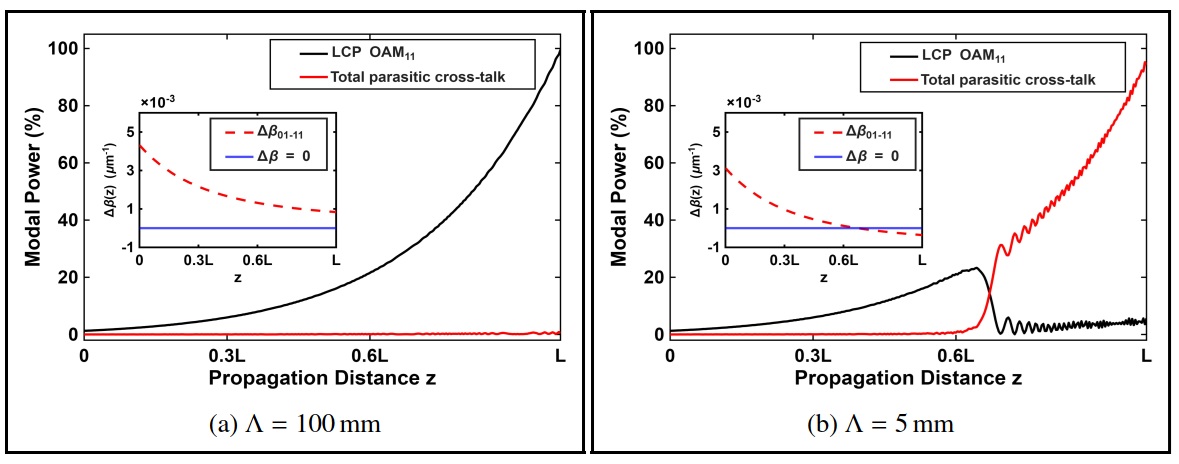}
 \caption{Intermodal power transfer from an input $\text{LCP OAM}_{11}$ to other guided modes. The coupling dynamics are compared for twist pitches of (a) $\Lambda=100\,\text{mm}$ and (b) $\Lambda=5\,\text{mm}$, under core offset perturbation ($d_{1}=0.3$) The insets illustrate phase mismatch evolution along the TSRCFA. The amplifier is counter-pumped from the wide side at $z=L$, and the modal power is normalized to the total output signal power. Simulation parameters are listed in Table~\ref{tab:sim_param}.}
    \label{fig:d1_OAM11_100mm_5mm_transfer}
\end{figure}

Transverse core offset is a common fabrication defect that can also be introduced purposely for controlled mode conversion. It is therefore important to analyze how this $N=1$ perturbation affects the evolution of guided modes in TSRCFAs. Setting $N=+1$ in Eq.~(\ref{selectionrule}) reveals that the perturbation couples modes satisfying $\Delta J=+1$ in TAM. Only modes from different azimuthal families meet this condition. According to Eq.~(\ref{twistedphasematching}), the phase matching condition for this perturbation differs from that of the elliptical perturbation. Figure~\ref{fig_core_offset_OAM11} illustrates the evolution of the full field of an input $\text{LCP OAM}_{11}$ (at $z=0$) signal mode propagating through the TSRCFA detailed in Sect.~$3.1$. The perturbation amplitude is set to $d_{1}=0.3$, and the corresponding twist pitches are set to $100\,\text{mm}$ (panel (a)) and $5\,\text{mm}$ (panel (b)).
For $\Lambda=100\,\text{mm}$, the calculated fidelity factor at four distinct propagation positions remains stably high as $100\,\%$, $\approx 99\,\%$, $\approx 99\,\%$, and $\approx 100\,\%$. A closer analysis of Fig.~\ref{fig_core_offset_OAM11} confirms that, unlike the elliptical deformation case, the injected $\text{LCP OAM}_{11}$ mode propagates almost robustly under this longer twist pitch. In contrast, for $\Lambda=5\,\text{mm}$ the fidelity factor evolves as $100\,\%$, $\approx 97\,\%$, $\approx 96\,\%$, and $\approx 5\,\%$. 
It shows a gradual initial decline, attributed to slight change in the polarization state, followed by a significant degradation after $z=0.6\,L$. The extreme reduction in fidelity at the output is driven primarily by the phase distribution (not the polarization state), which demonstrates an almost flat phase pattern not a helical one. 
The physical mechanism for this behavior is understood by the modal power exchange and phase matching condition shown in Fig.~\ref{fig:d1_OAM11_100mm_5mm_transfer}. 

\begin{figure}[htbp]
\centering
\includegraphics[width=11cm]{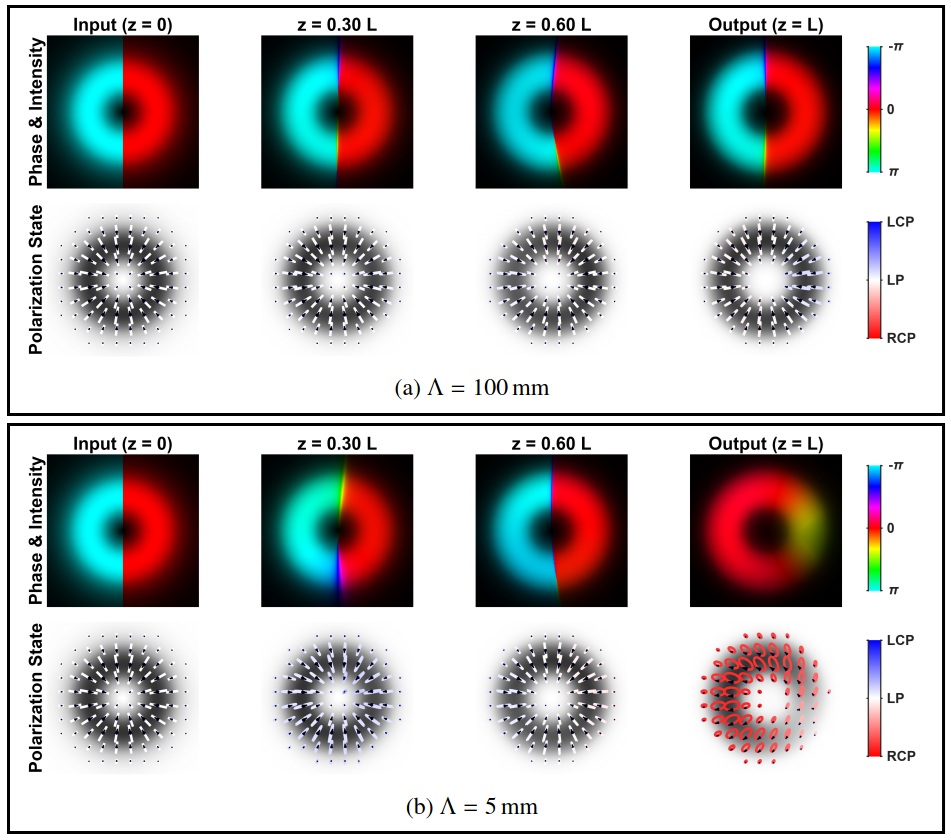}
  \caption{Evolution of the polarization, phase and intensity of an input $\text{TM}_{01}$ over propagation in the TSRCFA with twist pitches of (a) $100\,\text{mm}$ and (b) $5\,\text{mm}$, under core-offset perturbation ($d_{1}=0.3$). Simulation parameters are listed in Table~\ref{tab:sim_param}.}
    \label{fig_core_offset_TM01}
\end{figure}

For $\Lambda=100\,\text{mm}$ (panel (a)), there is almost no meaningful modal interactions, confirming the stable fidelity factor observed along the fiber. The inset depicts the longitudinally varying phase mismatch, $\Delta\beta(z)$, between input $\text{LCP OAM}_{11}$ and $\text{LCP OAM}_{01}$ (red dashed line) calculated using Eq.~(\ref{twistedphasematching}). Because this phase-mismatch curve never intersects the perfect phase-matching condition, $\Delta\beta=0$ (solid blue line), no efficient energy transfer occurs. However, the modal power exchange for $\Lambda=5\,\text{mm}$ (panel (b)) immediately explains the fidelity factor degradation. In this highly twisted regime, the longitudinal variation of the propagation constants drives the phase-matching curve across the zero-detuning line after $z=0.6\,L$, leading to a complete power transfer from the input mode. Simulation results confirms that $\text{LCP OAM}_{01}$ captures $\approx 94\%$ of the total output signal power, confirming the flat phase distribution observed at the fiber exit. Consequently, precise tuning of the twist pitch in a long geometrically perturbed ($N=1$ in this specific case) TSRCFA provide a mechanism for realizing a highly efficient $\text{OAM}_{01}\Leftrightarrow\text{OAM}_{11}$ mode converter. 
A corresponding analysis for an incident $\text{TM}_{01}$ modeunder identical simulation conditions is presented in Fig.~\ref{fig_core_offset_TM01}. 

The most physically profound difference between the $\text{OAM}_{11}$ and $\text{TM}_{01}$ excitation at $\Lambda=5\,\text{mm}$ lies in their output polarization states. Although the $l=0$ modal family captures the majority of the total output signal power in both scenarios, the $\text{TAM}$ selection rule defined in Eq.~(\ref{selectionrule}) dictates a dominant output LCP polarization state for the $\text{OAM}_{11}$ input (Fig.~\ref{fig_core_offset_OAM11}), while a highly output RCP polarization state is obtained for the $\text{TM}_{01}$ launch (Fig.~\ref{fig_core_offset_TM01}).
%\FloatBarrier
\subsection{Higher-order mode conversion}
\begin{figure}[htbp]
\centering
\includegraphics[width=11cm]{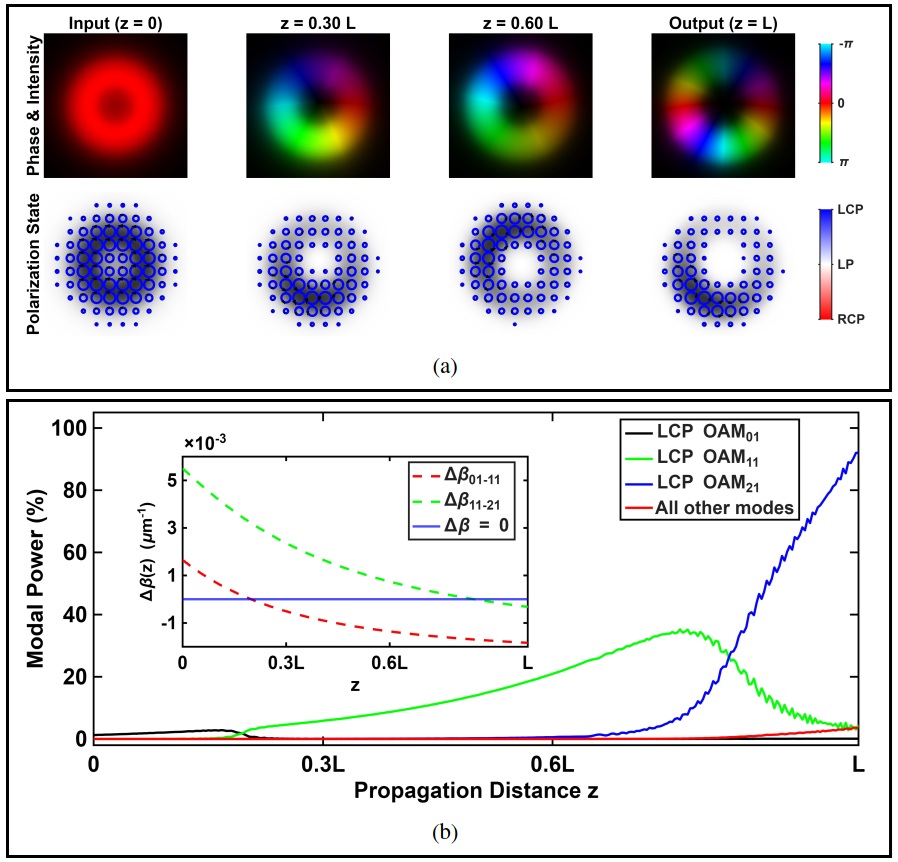}
   \caption{(a) Evolution of the polarization, phase and intensity of an input $\text{LCP OAM}_{01}$ over propagation in the perturbed TSRCFA with a twist pitch of $2.3\,\text{mm}$, under core-offset perturbation ($d_{1}=0.3$). (b) Corresponding modal power transfer. The inset shows the local phase mismatch for $\text{LCP (}\text{OAM}_{01}-\text{OAM}_{11}\text{)}$ (red dashed line) and $\text{LCP (}\text{OAM}_{11}-\text{OAM}_{21}\text{)}$ (green dashed line) pairs. Simulation parameters are listed in Table~\ref{tab:sim_param}.}
    \label{fig_HOMC_OAM01}
\end{figure}

Following the discussion on efficient $\text{OAM}$ mode conversion, we investigate the realization of higher-order $\text{OAM}$ mode converters in TSRCFAs by systematically adjusting the twist pitch of the structure. In Fig.~\ref{fig_HOMC_OAM01}, we analyze the evolution of the full vector field of an input $\text{LCP OAM}_{01}$ (at $z=0$) through the TSRCFA detailed in Table~\ref{tab:sim_param}, assuming a twist pitch of $\Lambda=2.3\,\text{mm}$ and a perturbation amplitude of $d_{1}=0.3$. 

A close inspection of the longitudinal phase dynamics in Fig.~\ref{fig_HOMC_OAM01} reveals that the input $\text{OAM}_{01}$ couples almost completely into $\text{OAM}_{11}$ around $z=0.3\,L$ (as confirmed by the modal exchange plot in the lower panel). However, at the output the field exhibits a helical phase structure corresponding to a topological charge of $l=2$. The power exchange dynamics confirm this cascaded conversion process, demonstrating that the $\text{LCP OAM}_{21}$ captures $\approx90\,\%$ of the total output signal power. Both the required twist pitch and the overall conversion efficiency can be adjusted by optimizing parameters like the ring thickness, numerical aperture of the ring guide, and the operating wavelength. 
\section{Conclusion}
In this work, we have investigated how different geometrical perturbations and twist pitches affect the evolution of the full-vector modal field, including its polarization, intensity and phase distributions, during propagation in TSRCFAs. We introduced a normalized complex modal-state fidelity factor, which provides a quantitative measure to evaluate any changes in the complete modal field. Our results show that a systematic choice of the twist pitch can ensure highly stable propagation of guided modes in the presence of structural perturbations. In addition to the stability analysis, we demonstrated that a typical twist pitch of $\Lambda=5\,\text{mm}$ can yield stable propagation of $\text{OAM}$ and $\text{TM}$ modes against an $N=2$ perturbation, while the same twist pitch provides a route to achieve highly efficient ($\approx 94\,\%$) OAM mode converters under an $N=1$ perturbation. Moreover, we highlighted a cascaded mode conversion process in which optical power transfers from an input $\text{OAM}_{01}$ to $\text{OAM}_{21}$ through an intermediate $\text{OAM}_{11}$. Although this study can be readily extended to evaluate the combined effect of different geometrical perturbations, analyzing each term separately provided a clear understanding of its individual role. These findings offer practical instructions for the design of $\text{OAM}$-based fiber amplifiers, $\text{OAM}$ mode converters and robust $\text{OAM}$ transmission in perturbed optical fibers.
% Force all remaining figures/tables to output before the end matter
\clearpage

% --- Funding & Acknowledgments (Native REVTeX syntax) ---
\begin{acknowledgments}
The authors acknowledge the financial support from the Research Council of Finland Flagship Programme (PREIN - decision Grant No. 320165) and the Horizon Europe EIC Pathfinder Open (101096317) project.
\end{acknowledgments}

% --- Disclosures & Data Availability (Standard unnumbered sections) ---
\section*{Disclosures}
The authors declare no conflicts of interest.

\section*{Data Availability}
Data underlying the results presented in this paper are not publicly available at this time but may be obtained from the authors upon reasonable request.

% --- References ---
\bibliography{sample}

\end{document}